\documentclass{svmult}

\usepackage{makeidx}         

\usepackage{graphicx}        

\usepackage{multicol}        

\makeindex             

\begin{document}

\title*{Origins of Hydrodynamics for a Granular Gas}



\author{James W. Dufty\inst{1}\and J. Javier Brey\inst{2}}



\institute{Department of Physics, University of Florida,
Gainesville, FL 32611, USA  \texttt{dufty@phys.ufl.edu} \and
Fisica Teorica, Universidad de Sevilla, Apartado de Correos 1065,
E-41080 Sevilla, Spain \texttt{brey@cica.es}}

%




%

\maketitle

\begin{abstract}

The basis for a hydrodynamic description of granular gases is
discussed for a low density gas of smooth, inelastic hard spheres.
The more fundamental mesoscopic description is taken to be the
nonlinear Boltzmann kinetic equation. Two arguments are presented
in favor of a hydrodynamics for granular gases. The first one is
the concept of a ''normal'' solution and its explicit approximate
construction via the Chapman-Enskog method. The second is the
demonstration of hydrodynamic modes in the spectrum of the
generator for the dynamics of small spatial perturbations of the
homogeneous reference state. In the first case, a derivation of
the nonlinear hydrodynamic equations is given to Navier-Stokes
order, with explicit expressions for the transport coefficients.
The approach is formal and the context of the derivation is left
unspecified, although internal mathematically consistency is
demonstrated. The second method is more restricted, leading only
to linearized hydrodynamics, but with the potential to define more
sharply the context of hydrodynamics.

\end{abstract}

\section{Introduction}

\label{sec:1} The form and validity of a hydrodynamic description
for granular gases remains a controversial subject \cite{G01}. The
objective here is to formulate the conceptual and mathematical
questions in terms of the underlying Boltzmann kinetic equation,
and to describe some recent results providing support for
hydrodynamics \cite{BDR03,DB03,DB04}. Two distinct approaches will
be discussed. The first is a consideration of ``constitutive
equations'' for a closure of the exact macroscopic balance
equations for the average mass, energy (or temperature), and
momentum (or flow velocity). These equations follow directly from
the existence of a ``normal'' solution to the Boltzmann equation.
The expected conditions for a normal solution, e.g., away from
spatial or temporal boundary layers, characterize the context in
which a hydrodynamic description can be expected. The explicit
construction of an approximate normal solution under conditions of
small spatial variation of the hydrodynamic fields is given by an
extension of the Chapman-Enskog method \cite{BDKS99}. It is
emphasized that no additional assumptions beyond the normal form
and small gradients are required by the inelasticity of
collisions. Differences from normal gases (e.g., the reference
state, linear stability, form of the constitutive equations) are a
consequence of the method, not a limitation on it. The resulting
hydrodynamics comprises the Navier-Stokes approximation for a
granular gas. Finally, it is emphasized that a breakdown of the
latter does not imply the absence of a hydrodynamic description.
Indeed, it is an interesting new challenge of granular gases to
obtain constitutive equations for more general conditions
\cite{G01,SGD03}.

The second means to explore the existence and context of
hydrodynamics is to consider small spatial perturbations of a
strictly homogeneous state. This problem is somewhat simpler as it
is governed by the linearized Boltzmann equation, and it is then
sufficient to consider a single Fourier component. The spectrum of
the generator for the dynamics determines all possible
excitations, and it can be asked if the hydrodynamic excitations
are among these. For consistency, the hydrodynamic excitations
found in this way must be the same as those from the Navier-Stokes
equations under the same conditions.

Recently, the eigenvalue problem for this generator has been
considered \cite{BDR03,DB03,DB04}. There is a technical problem
associated with the ``cooling'' of the reference homogeneous
state, which is controlled by the introduction of suitable
dimensionless variables. This entails a corresponding change in
the form of the generator for the dynamics. The existence of
hydrodynamic modes has been proved by the exact construction of
the hydrodynamic eigenvalues and eigenfunctions in the long
wavelength limit. Their form to Navier-Stokes order is then
obtained by perturbation theory, and complete agreement with the
results from the Chapman-Enskog method is found. The dominance of
these hydrodynamic excitations at long times requires that their
eigenvalues be the smallest ones and be isolated from the rest of
the spectrum. Some comments are offered on why this is to be
expected. Elsewhere in this volume \cite{BD04} these expectations
are confirmed for an exactly solvable kinetic model.

The objective of this brief overview is to clarify the concept of
a hydrodynamic description in its most general sense, and to
sharpen the sufficient conditions for its validity.


\section{Boltzmann Equation and Balance Equations}
\label{sec2}

Consider a system of $N$ smooth hard spheres of mass $m$ and
diameter $\sigma $ in a large volume $V$. If the density $n=N/V$
is sufficiently small, $n\sigma ^{3}<<1$, the phase space density,
$f(\mathbf{r},\mathbf{v},t)$, for the number of particles with
position $\mathbf{r}$ and velocity $\mathbf{v}$ at time $t$ is
determined from the Boltzmann equation \cite{GS95,D01,vNE01},
\begin{equation}
\left( \frac{\partial }{\partial t}+\mathbf{v}\cdot \nabla \right)
f(\mathbf{ r},\mathbf{v},t)=J\left[
\mathbf{r},\mathbf{v}|f(t)\right] , \label{2.1}
\end{equation}
where $J$ is the inelastic Boltzmann collision operator. The most
important properties following from this collision operator for
the following discussion are:
\begin{equation} \int
d\mathbf{v}\left(
\begin{array}{c}
1 \\
\mathbf{v} \\
\frac{1}{2}m\left( \mathbf{v}-\mathbf{u}\right) ^{2}
\end{array}
\right) J\left[ \mathbf{r},\mathbf{v}|f(t)\right] =\left(
\begin{array}{c}
0 \\
\mathbf{0} \\
-\frac{3}{2}n(\mathbf{r},t)T(\mathbf{r},t)\zeta (\mathbf{r},t)
\end{array}
\right) .  \label{2.4}
\end{equation}
The two zeros on the right side of (\ref{2.4}) follow from
conservation of mass and momentum. The last term results from
non-conservation of energy, where $\zeta $ is the fractional rate
of decrease in the energy due to the inelasticity,
\begin{equation}
\zeta (\mathbf{r},t)=\frac{(1-\alpha^{2}) m\pi \sigma ^{2}}{24n(
\mathbf{r},t)T(\mathbf{r},t)}\int \,d\mathbf{v}\,\int
\,d\mathbf{v} _{1}\,g^{3}\
f(\mathbf{r},\mathbf{v},t,)f(\mathbf{r},\mathbf{v}_{1},t).
\label{2.5}
\end{equation}
Here $ \mathbf{g}= \mathbf{v}-\mathbf{v}_{1}$ and $\alpha $ is the
restitution coefficient ($\alpha =1$ is the elastic limit). The
``hydrodynamic fields'' appearing in (\ref{2.4}) and (\ref{2.5})
are the density $n(\mathbf{r},t)$, the temperature
$T(\mathbf{r},t)$, and the macroscopic flow velocity
$\mathbf{u}(\mathbf{r},t)$. These are defined in terms of velocity
moments of $f(\mathbf{r},\mathbf{v},t)$ by
\begin{equation}
\left(
\begin{array}{c}
n(\mathbf{r},t) \\
n(\mathbf{r},t)\mathbf{u}(\mathbf{r},t) \\
\frac{3}{2}n(\mathbf{r},t)T(\mathbf{r},t)
\end{array}
\right) =\int d\mathbf{v}\left(
\begin{array}{c}
1 \\
\mathbf{v} \\
\frac{1}{2}m\left( \mathbf{v}-\mathbf{u}\right) ^{2}
\end{array}
\right) f(\mathbf{r},\mathbf{v,}t).  \label{2.4a}
\end{equation}

The macroscopic balance equations for these fields follow directly
from the Boltzmann equation using the properties (\ref{2.4}),
\begin{equation}
\partial _{t}n+\nabla \cdot \left( n\mathbf{u}\right) =0,  \label{2.5a}
\end{equation}
\begin{equation}
\left( \partial _{t}+\mathbf{u}\cdot \nabla \right)
u_{i}+(mn)^{-1}{\nabla}_{j}P_{ij}=0,  \label{2.5b}
\end{equation}
\begin{equation}
\left( \partial _{t}+\mathbf{u}\cdot \nabla +\zeta \right)
T+\frac{2}{3n} \left( P_{ij}{\nabla }_{j}u_{i}+\nabla \cdot
\mathbf{q}\right) =0. \label{2.5c}
\end{equation}
The pressure tensor $P_{ij}$ and heat flux $\mathbf{q}$ are
defined by
\begin{equation}
P_{ij}(\mathbf{r},t)=\int d\mathbf{v}\,m (v_{i}
-u_{i})(v_{j}-u_{j})f(\mathbf{r},\mathbf{v},t), \label{2.5d}
\end{equation}
\begin{equation}
\mathbf{q}(\mathbf{r},t)=\int
d\mathbf{v}\frac{m}{2}(\mathbf{v}-\mathbf{u})^{2}
(\mathbf{v}-\mathbf{u}) f(\mathbf{r},\mathbf{v},t). \label{2.5e}
\end{equation}
In the above equations and in the following, summation over
repeated indexes is implicit. For a normal gas with elastic
collisions, Eqs. (\ref{2.5a})-(\ref{2.5e}) are the local
macroscopic conservation laws for mass, momentum, and energy. The
only formal difference here is the appearance of the ``cooling
rate'', $ \zeta $, in the equation for the temperature. Moreover,
all the explicit dependence on the restitution coefficient
$\alpha$ occurs in this cooling rate.

The macroscopic balance equations are the starting point for
developing a hydrodynamic description of the gas. They are an
exact consequence of the Boltzmann equation, but are of limited
utility without further specification of the three unknown
functions $P_{ij}$, $\mathbf{q}$, and $\zeta$. Since these are
specific functionals of $f(\mathbf{r},\mathbf{v},t)$, the above
requires a solution to the Boltzmann equation. The special class
of solutions leading to hydrodynamics, normal solutions, are
described in the next section.

\section{Normal Solution, Constitutive Equations, and Hydrodynamics}
\label{s3}

A sufficient condition for a hydrodynamic description is the
existence of a normal solution to the Boltzmann equation. To
describe it, denote the hydrodynamic fields collectively by the
set $\left\{ y_{\alpha }(\mathbf{r} ,t)\right\} $. Then, a normal
solution is defined as one for which all space and time dependence
of the distribution function is determined by the hydrodynamic
fields, i.e.,
\begin{equation}
f(\mathbf{r},\mathbf{v},t)=f\left[ \mathbf{r},\mathbf{v},t |
\left\{ y_{\alpha }\right\} \right],  \label{3.1}
\end{equation}
such that
\begin{equation}
\left(
\begin{array}{c}
\partial _{t} \\
\nabla
\end{array}
\right) f \left[ \mathbf{r},\mathbf{v},t | \left\{ y_{\alpha
}\right\} \right]=\int dt^{\prime } \int d\mathbf{r}^{\prime
}\frac{\delta f(\mathbf{r},\mathbf{v},t | \left\{ y_{\alpha
}\right\} )}{\delta y_{\beta }(\mathbf{r}^{\prime },t^{\prime
})}\left(
\begin{array}{c}
\partial _{t^{\prime }} \\
\nabla ^{\prime }
\end{array}
\right) y_{\beta }(\mathbf{r}^{\prime },t^{\prime }). \label{3.2}
\end{equation}
If such a solution can be found, its use in the functionals
(\ref{2.5}, (\ref{2.5d}), and (\ref{2.5e}), converts them to
normal forms as well,
\begin{equation}
\zeta=\zeta \left[ \mathbf{r} ,t \mid \left\{ y_{\alpha
}\right\}\right], \quad P_{ij}=P_{ij}\left[ \mathbf{r},t\mid
\left\{ y_{\alpha }\right\}\right],  \quad \mathbf{q}=\mathbf{q}
\left[\mathbf{r},t\mid \left\{ y_{\alpha }\right\}\right].
\label{3.3}
\end{equation}
Such normal forms are called ``constitutive'' equations. Finally,
use of these normal forms in the macroscopic balance equations
gives a {\em{closed}} set of equations for the hydrodynamic
fields:
\begin{equation}
\partial _{t}y_{\alpha }(\mathbf{r},t)=N_{\alpha }\left[ \mathbf{r},t\mid
\left\{ y_{\beta }\right\} \right] ,  \label{3.4}
\end{equation}
where $N_{\alpha }\left[ \mathbf{r},t\mid \left\{ y_{\beta
}\right\} \right] $ is, in general, a non-local, non-linear
functional of the set of fields $ \left\{ y_{\beta }\right\} $.
Equations (\ref{3.4}) are the hydrodynamic equations in their most
general form, encompassing both rheological and viscoelastic
nonlinear transport. Such a general context is typically reserved
for complex fluids (e.g., colloids, large molecular weight fluids)
and is of little interest for normal gases. However, structurally
simple granular gases can exhibit many of the non-Newtonian
properties of complex fluids. Therefore, in discussing the
applicability of a hydrodynamic description for granular gases, it
is important to admit the possibility of a closed set of equations
more complex than the local partial differential equations of the
Navier-Stokes approximation (see below).

The construction of a hydrodynamic description is seen to be
intimately connected to finding a normal solution to the Boltzmann
equation. In fact the two problems comprise a single
self-consistent problem. For a normal solution, the Boltzmann
equation (\ref{2.1}) becomes
\begin{eqnarray}
\int dt^{\prime } \int d\mathbf{r}^{\prime }\frac{\delta f\left[
\mathbf{r},\mathbf{v} ,t\mid \left\{ y_{\alpha }\right\}
\right]}{\delta y_{\beta }(\mathbf{r}^{\prime },t^{\prime })}
\Big( N_{\beta }\left[ \mathbf{r}^{\prime},t^{\prime }\mid \left\{
y_{\gamma} \right\} \right]  &+&  \mathbf{v}\cdot \nabla ^{\prime
}y_{\beta } (\mathbf{r}^{\prime},t^{\prime}) \Big) \nonumber \\
& = & J\left[ \mathbf{r},\mathbf{v}|f \right] . \label{3.5}
\end{eqnarray}
For specified fields, this is an equation for the velocity
dependence of the normal distribution as a functional of the
fields. This dependence then allows determination of the normal
forms in (\ref {3.3}), and hence $N_{\alpha }\left[
\mathbf{r},t\mid \left\{ y_{\beta }\right\} \right] $. Finally,
solution of Eq.\ (\ref{3.4}), with suitable initial and boundary
conditions, provides the explicit forms for the fields, and
completes the normal solution. The existence and determination of
this solution is the central problem for establishing a
hydrodynamic description for both normal and granular gases. The
concept of a normal solution and its use in the macroscopic
balance equations makes no special reference to the possible
inelasticity of collisions.

To clarify the conditions under which a normal solution could be
expected, consider first a normal gas with elastic collisions in
an initial non-equilibrium state with specified hydrodynamic
fields $\left\{ y_{\alpha }\left( \mathbf{r} ,t=0\right) \right\}
$,  whose values vary smoothly across the system. In each small
region (e.g., cells of $20$-$30$ particles), the velocity
distribution approaches a Maxwellian characterized by the
hydrodynamic fields at its central point $\mathbf{r}$.
Subsequently, exchange of mass, energy, and momentum tends to
equilibrate these fields to uniform values (or to steady values if
the system is driven). The first stage, approach to a universal
form for the velocity distribution, occurs after a few collisions.
This establishes the normal form of the solution, where the
hydrodynamic fields and their gradients characterize the state.
Deviations from the Maxwellian are due to fluxes of mass,
momentum, and energy across the cells. These fluxes are
proportional to the differences in values of the fields (i.e., to
their spatial gradients). The second stage is the evolution of the
distribution through the changing values of the fields, according
to the hydrodynamic equations.

This two-stage evolution can be expected for granular gases as
well. The initial velocity relaxation will not approach the local
Maxwellian, but some other corresponding normal distribution
determined from the inelastic Boltzmann equation (see below).
Subsequently, the deviations from this normal distribution
characterizing spatial  inhomogeneities,  will again be via the
macroscopic balance equations for the granular gas.

\subsection{Constructive Example: Navier-Stokes Approximation}

It is easily verified that the normal solution cannot be simply a
local function of the fields alone, i.e. $f\neq f[\mathbf{v}\mid
\left\{ y_{\alpha }\right\} ]$. However, the nonlocal functional
dependence can be converted to a local function of the fields and
all their gradients,
\begin{equation}
f=f\left( \mathbf{v},\left\{ y_{\alpha }\left( \mathbf{r},t\right)
\right\} ,\left\{ \nabla y_{\alpha }\left( \mathbf{r},t\right)
\right\} ,\left\{ \nabla \nabla y_{\alpha }\left(
\mathbf{r},t\right) \right\} , \ldots \right) .  \label{3.6}
\end{equation}
The time derivatives have not been included, since they can be
expressed formally in terms of the gradients through the
hydrodynamic equations (\ref {3.4}), with the corresponding local
form $N_{\alpha }\left( \left\{ y_{\beta }\left(
\mathbf{r},t\right) \right\} ,\left\{ \nabla y_{\beta }\left(
\mathbf{r},t\right) \right\} ,\left\{ \nabla \nabla y_{\beta
}\left( \mathbf{r},t\right) \right\} , \ldots \right)$. Such a
representation is useful when some higher order gradients can be
neglected. For example, if the initial and boundary conditions
suggest nearly linear hydrodynamic profiles (such as for shear
flow between parallel moving plates), the above can be estimated
as $ f \sim $\ $f\left( \mathbf{v},\left\{ y_{\alpha }\left(
\mathbf{r}, t\right) \right\} ,\left\{ \nabla y_{\alpha }\left(
\mathbf{r},t\right) \right\} \right) $. This approximation could
still include large gradients and nonlinear transport.

A simpler case is that for which all the spatial gradients are
small. More precisely, consider states for which the fractional
changes in the hydrodynamic fields are small over a mean free
$\ell $, i.e., $\ell \left| \nabla y_{\alpha }\left(
\mathbf{r},t\right) /y_{\alpha }\left( \mathbf{r},t\right) \right|
\sim \epsilon <<1$ and $ \ell \left| \nabla ^{n+1}y_{\alpha
}\left( \mathbf{r},t\right) /\nabla ^{n}y_{\alpha }\left(
\mathbf{r},t\right) \right| \sim \epsilon <<1$. The parameter
$\epsilon $ and is called the  \textit{uniformity parameter}. The
normal solution can now be calculated perturbatively as an
expansion in this small parameter. To first order in $\epsilon $,
it must be linear in the hydrodynamic gradients
\begin{equation}
f(\mathbf{r},\mathbf{v},t) \sim f^{(0)}\big( \mathbf{v},\left\{
y_{\alpha }\left( \mathbf{r},t\right) \right\} \big) \big(
1+\epsilon \mathbf{X}_{\beta }\left[ \mathbf{v},\left\{ y_{\gamma
}\left( \mathbf{r},t\right) \right\} \right]\cdot \nabla y_{\beta
}\left( \mathbf{r},t\right) \big) .  \label{3.6b}
\end{equation}
Before discussing the specific forms for $f^{(0)}$ and
$\mathbf{X}_{\alpha }$, the normal forms for  $\zeta
(\mathbf{r},t,)$, $P_{ij}(\mathbf{r},t,)$, and
$\mathbf{q}(\mathbf{r},t,)$ can be determined directly by use of
(\ref{3.6b}) in the functionals (\ref {2.5}), (\ref{2.5d}), and
(\ref{2.5e}), and invoking the fluid symmetry,
\begin{equation}
\zeta=\left( 1-\alpha ^{2}\right) n\left( \mathbf{r} ,t\right)
\sigma ^{2}\left( \frac{T\left( \mathbf{r},t\right) }{m}
\right)^{1/2}\overline{ \zeta },  \label{3.7}
\end{equation}
\begin{equation}
\mathbf{q} =-\kappa \left[ T\left( \mathbf{r} ,t\right) \right]
\nabla T\left( \mathbf{r},t\right) -\mu \left[ T\left(
\mathbf{r},t\right) ,n\left( \mathbf{r},t\right) \right] \nabla
n\left( \mathbf{r},t\right),   \label{3.8}
\end{equation}
\begin{eqnarray}
P_{ij} & = & p\left( \mathbf{r},t \right) \delta _{ij} -\eta
\left[ T\left( \mathbf{r},t\right) \right] \bigg[
\nabla_{j}u_{i}\left( \mathbf{r},t\right)  +\nabla_{i}u_{j}\left(
\mathbf{r},t \right)
  \nonumber \\
&& - \left. \frac{2}{3}\delta _{ij}\nabla \cdot \mathbf{u}\left(
\mathbf{r},t\right) \right] . \label{3.9}
\end{eqnarray}
Here $p=nT$ is the pressure, $\overline{\zeta }$ is a
dimensionless constant determined from $f^{(0)} $, and the
transport coefficients $\kappa $, $\mu $, and $\eta $ are linear
functionals of the $\mathbf{X}_{\alpha }$. Their exact expressions
in terms of $f^{(0)}$ and the $\mathbf{X} _{\alpha }$ are given
elsewhere \cite{BDKS99,GD99} and will not be repeated here. In the
expression for the cooling rate (\ref{3.7}), only the leading
contribution (zeroth order in the gradients) is given. In
principle, consistency would require to consider up to second
order in the gradients, since this is the order of the terms
associated with the heat flux and the pressure tensor.
Nevertheless, contributions from the cooling rate of first order
in the gradients of the fields are seen to vanish because of
symmetry reasons, and those of second order have been shown to be
negligible as compared with the corresponding contributions from
Eqs.\ {\ref{3.8}) and (\ref{3.9}) \cite{BDKS99}.

The macroscopic balance equations of the last section together
with (\ref {3.7})-(\ref{3.9}) constitute the Navier-Stokes
hydrodynamic equations. Their form is the same as for a normal
fluid, except for the presence of the cooling rate $\zeta $ and
the new transport coefficient $\mu $ in the equation for the heat
flux. Of course, the values of the transport coefficients are
different, depending on the value of the restitution coefficient
$\alpha$.

The leading term $f^{(0)}$ in  (\ref{3.6b}) is determined from the
Boltzmann equation to order $\epsilon ^{0}$,
\begin{equation}
-\zeta  T\frac{\partial f^{(0)}}{\partial T}=J\left[
\mathbf{r},\mathbf{ v}|f^{(0)}\right] .  \label{3.13}
\end{equation}
Since $f^{(0)}$ is normal, it follows from dimensional analysis
that its dependence on temperature can occur only through
normalization and a scaling of the velocity. Then (\ref{3.13})
becomes:
\begin{equation} \frac{\zeta}{2}\,  \frac{\partial}{\partial
\mathbf{v}} \cdot \left[ \left( \mathbf{v}-\mathbf{u }\right)
f^{(0)}\right] =J\left[ \mathbf{r},\mathbf{v}|f^{(0)}\right] .
\label{3.13a}
\end{equation}
This equation also arises when considering the Boltzmann equation
for a spatially homogeneous state, the so-called homogeneous
cooling state (HCS), as discussed in the next section. Here, it is
determining the same velocity dependence, but scaled with respect
to the exact hydrodynamic fields for the nonuniform, nonstationary
state. Thus the solution here is more appropriately named the
\textit{local} HCS. It is important to note that this leading
order ``reference'' state is determined by the Chapman-Enskog
procedure itself, not by an arbitrary, independent choice.

Similarly, the first corrections, determined by
$\mathbf{X}_{\alpha }$, are obtained from the Boltzmann equation
to order $\epsilon $, with the result:
\begin{eqnarray}
\bigg[ L \mathbf{X}_{\alpha } & - & \left.f^{(0)-1} \zeta T\,
\frac{\partial}{\partial T} \left( f^{(0)}\mathbf{X}_{\alpha
}\right) \right] \cdot \nabla y_{\alpha
} -  \mathbf{X}_{T} \cdot \nabla \left( \zeta T\right)   \nonumber \\
&=&-\frac{\partial \ln f^{(0)}}{\partial y_{\alpha }}\left[
N_{\alpha }^{(1)}\left( \mathbf{r},t\mid \left\{ y_{\beta
}\right\} \right) +\mathbf{v }\cdot \nabla y_{\alpha}\right] ,
\label{3.14}
\end{eqnarray}
where $N_{\alpha}^{(1)}$ is the term of order $\epsilon$ in the
expansion of $N_{\alpha}$, $\mathbf{X}_{T}$ is the coefficient of
$\nabla T$ in Eq. (\ref{3.6b}), and $L$ is the linearized
Boltzmann collision operator,
\begin{equation}
LY (\mathbf{v})=-f^{(0)-1}(\mathbf{v}) \int d\mathbf{v}^{\prime
}\, \frac{\delta J\left[ \mathbf{r}, \mathbf{v}|f^{(0)}\right]
}{\delta f^{(0)}(\mathbf{v}^{\prime },\left\{ y_{\alpha }\right\}
)}f^{(0)}(\mathbf{v}^{\prime})Y(\mathbf{v}^{\prime }),
\label{3.15}
\end{equation}
for arbitrary $Y(\mathbf{v})$. By equating coefficients of the
hydrodynamic gradients in (\ref{3.14}), the coefficients having
independent scalar, vector, and tensor transformation properties,
a set of linear inhomogeneous integral equations are obtained. The
Fredholm conditions have been proved for these equations, so that
existence of solutions is assured. The details are described in
\cite{BDKS99} and \cite{GD99}. In summary, a normal solution to
the Boltzmann equation up through first order in $\epsilon$ is
given by (\ref{3.6b}). This solution determines the constitutive
equations in normal form also to this order. These constitutive
equations together with the macroscopic balance equations, give
approximate hydrodynamic equations which are the Navier-Stokes
equations for a granular gas. The transport coefficients in these
equations are also provided by the derivation.

The main point of the discussion in this section has been to show
that hydrodynamic equations for granular gases have the same
conceptual basis as for normal gases: 1) exact macroscopic balance
equations for $n$, $T$, and $\mathbf{u}$, with exact expressions
for the unknown fluxes (and the cooling rate) as functionals of
the distribution function $f$; 2) a rapid relaxation of $f$ to a
normal solution. These two ingredients are sufficient for the
existence of a hydrodynamic description (closed, deterministic
equations for the hydrodynamic fields). The construction of the
normal solution is a difficult, unsolved problem in general.
However, for states with small spatial gradients, the
Chapman-Enskog perturbation expansion leads to an explicit
approximate normal solution. No explicit reference to inelasticity
occurs at this level. The detailed computation of the normal
solution and characterization of the space and time scales for its
validity does, of course, entail differences from that for normal
gases. But these are technical issues that should be separated
from the conceptual basis.

Property 2) above is the most uncertain component of the
discussion. If a normal solution exists, what is its context? For
molecular gases, the assumption is that it applies only outside
certain ``boundary layers'': a few mean free paths away from
spatial boundaries, a few mean free times later than any imposed
initial conditions, and away from ``discontinuities'' (e.g., shock
fronts). Similar restrictions are expected to apply for granular
gases, and it may be that these are stronger constraints in that
case. In addition, there is a new time scale for granular gases
set by the cooling rate. The hydrodynamic equations described here
include the dynamics of cooling, so there are new questions
regarding the separation of hydrodynamic from all other time
scales. Monte Carlo and molecular dynamics simulations suggest
that the separation remains valid even at strong dissipation, but
it remains to demonstrate this directly from the Boltzmann
equation. The next section provides a more restricted context in
which this question can be posed clearly.

\section{Linearized Boltzmann Equation and Hydrodynamic Modes}
\label{s4}

In this section, the existence of hydrodynamics for granular gases
is considered from a different perspective, independent of the
concept of a normal solution to the Boltzmann equation. The idea
is to study the initial value problem for small perturbations of
the homogeneous state of an isolated system. In this case, the
Boltzmann equation can be linearized about the homogeneous
reference state. For appropriate dimensionless variables, the
general solution is characterized by a linear eigenvalue problem.
The eigenvalues determine all modes of excitation in the gas due
to the small perturbations. If hydrodynamics exists, the
hydrodynamic modes must be among these excitations. These modes
are defined in the long wavelength limit as being the same as
those obtained from the exact linearized macroscopic equations The
reference homogeneous state for an isolated granular gas is taken
of the form
\begin{equation}
f_{hcs}\left( \mathbf{v}, t\right) =nv_{0}^{-3}(t)f_{hcs}^{\ast
}\left[ v/v_{0}(t)\right], \label{4.1}
\end{equation}
with
\begin{equation}
v_{0}(t)=\left( \frac{2T(t)}{m} \right)^{1/2}. \label{4.1a}
\end{equation}
Use of this scaling form in the expression for the cooling rate
(\ref{2.5}), gives again the form (\ref{3.7}), now as an exact
result,
\begin{eqnarray}
\zeta (t) & = & \frac{(1-\alpha)^{2} \pi \sigma^{2} n}{12}\,
\left( \frac{2 T(t)}{m} \right)^{1/2} \nonumber \\
& & \times \int \,d\mathbf{v}\,\int \, d \mathbf{v}_{1}\, |
\mathbf{v}-\mathbf{v}_{1}| ^{3}\ f_{hcs}^{\ast }\left( v\right)
f_{hcs}^{\ast }\left( v_{1}\right) . \label{4.2}
\end{eqnarray}

The macroscopic balance equations for the homogeneous state reduce
to $ \partial _{t}n=\partial _{t}\mathbf{u}=0$, and
\begin{equation}
\partial _{t}\ln T(t)=- \zeta (t).  \label{4.3}
\end{equation}
Since the dependence of $\zeta (t)$ on the temperature is known,
the time dependence of the temperature can be calculated.
Substitution of (\ref{4.1}) into the Boltzmann equation gives the
equation for $f_{hcs}$. It is found to be the same as that for the
local HCS distribution $f^{(0)}$, given by (\ref{3.13}) or
(\ref{3.13a}) above. The difference here is that the hydrodynamic
fields parameterizing this distribution are those for the
homogeneous state, whereas in the Chapman-Enskog context they are
the fields for a general inhomogeneous state.  As already
mentioned, the distribution $f_{hcs}$ is referred to as the HCS
solution. Its velocity dependence is the same as that of the
\textit{local} HCS solution.

In the following, it will be assumed that $f_{hcs}$ is known.
Consider an initial \emph{small }spatial perturbation of the HCS
(other examples might arise from small external forces or boundary
conditions). Since the perturbation is small initially, it is
assumed that the deviation of the distribution function from
$f_{hcs}$ remains small over the times of interest, such that the
Boltzmann equation can be linearized. Consequently, it is
sufficient to consider a single Fourier component,
\begin{equation}
f\left( \mathbf{r},\mathbf{v},t\right) =f_{hcs}\left(
\mathbf{v},t\right) \left[ 1+e^{i\mathbf{k\cdot r}}g\left(
\mathbf{k,v},t\right) \right]. \label{4.4}
\end{equation}
Substitution into the Boltzmann equation and retaining only linear
order in  $g$ gives the linear Boltzmann equation,
\begin{equation}
\left[ \partial _{t}+i\mathbf{k\cdot
v+}L\mathbf{+}f_{hcs}^{-1}(\mathbf{v}) \frac{\zeta}{2}
\frac{\partial}{\partial \mathbf{v}} \cdot \left(
\mathbf{v}f_{hcs}\right) \right] g (\mathbf{k},\mathbf{v},t)=0.
\label{4.5}
\end{equation}
Here, $L$ is the linear operator defined in (\ref{3.15}) evaluated
at $ f^{(0)}\rightarrow f_{hcs}$.

The time dependence of the reference HCS leads to an explicit time
dependence of $L$,  $\zeta $, and $f_{hcs}$. This can be removed
by introducing the following dimensionless variables
\begin{equation}
s=\int_{0}^{t}dt^{\prime }\frac{v_{0}(t^{\prime})}{\ell }, \quad
\quad \mathbf{v} ^{\ast }=\frac{\mathbf{v}}{v_{0}(t)}, \quad \quad
\mathbf{k}^{\ast }= \mathbf{k}\ell,   \label{4.6}
\end{equation}
\begin{equation}
g^{\ast }\left( \mathbf{k}^{\ast },\mathbf{v}^{\ast },s\right)
=g\left( \mathbf{k},\mathbf{v},t\right) , \quad \quad \zeta^{\ast
}=\frac{\ell \zeta }{ v_{0}(t)},  \label{4.7}
\end{equation}
where $\ell =1/n\sigma ^{3}$ is proportional to the mean free path
and $s$ is proportional to the average number of collisions per
particle in the time interval $\left( 0,t\right) $. The linearized
Boltzmann equation then becomes
\begin{equation}
\left( \partial _{s}+i\mathbf{k}^{\ast }\mathbf{\cdot v}^{\ast
}+\Lambda ^{\ast }\right) g^{\ast }=0,  \label{4.8}
\end{equation}
where the linear collision operator $\Lambda^{*} $ is defined by
\begin{equation}
\Lambda ^{\ast }Y=L^{\ast }Y+\frac{\zeta^{*}}{2}\, f_{hcs}^{\ast
-1} \frac{\partial}{\partial \mathbf{v}} \cdot \left(
\mathbf{v}f_{hcs}^{\ast }Y \right), \quad \quad L^{\ast
}=\frac{\ell }{v_{0}(t)}\, L,  \label{4.9}
\end{equation}
It is seen that the operator $\Lambda ^{\ast }$ is not simply the
linearized Boltzmann collision operator in dimensionless form, but
it includes as well an operator that describes the effects of
cooling explicitly.

Normalization of the distribution function requires only that
$f_{hcs}^{\ast }g^{\ast }$ be integrable. However, we make the
stronger requirement that $g^{\ast }$ be an element of a Hilbert
space with scalar product
\begin{equation}
\left( a,b\right) \equiv \int d\mathbf{v}^{\ast }f_{hcs}^{\ast
}\left( v^{\ast }\right) a^{+}(\mathbf{v}^{\ast
})b(\mathbf{v}^{\ast }), \label{4.10}
\end{equation}
where $a^{+}$ denotes the complex conjugated of $a$. The formal
solution to the linear Boltzmann equation is
\begin{equation}
g^{\ast }\left( \mathbf{k}^{\ast },\mathbf{v}^{\ast },s\right) =
\oint \frac{ dz}{2\pi i}e^{-zs}\mathcal{R}(z)g^{\ast }\left(
\mathbf{k}^{\ast },\mathbf{v} ^{\ast },0\right), \label{4.11}
\end{equation}
\begin{equation}
\mathcal{R}(z)=\left( z-i\mathbf{k}^{\ast } \mathbf{\cdot v}^{\ast
}-\Lambda^{*} \right) ^{-1},  \label{4.11a}
\end{equation}
where the contour encloses the entire spectrum of
$i\mathbf{k}^{\ast } \mathbf{\cdot v}^{\ast }+\Lambda^{*} $ , both
point and residual, counterclockwise. All linear excitations are
determined from this spectrum. The hydrodynamic modes, whenever
they exist, are part of the point spectrum so it is sufficient to
study the eigenvalue problem
\begin{equation}
\left( i\mathbf{k}^{\ast }\mathbf{\cdot v}^{\ast }+\Lambda^{*}
\right) \psi _{n}( \mathbf{v}^{\ast })=\lambda _{n}(k^{*})\psi
_{n}(\mathbf{v}^{\ast }). \label{4.12}
\end{equation}

The hydrodynamic modes are defined from the macroscopic balance
equations in the following way. First, a normal form for the
cooling rate, pressure tensor, and heat flux is assumed to give a
closed set of equations (hydrodynamics). Next, these equations are
linearized about the HCS state ($ n, \mathbf{v}=0, T(t)$). The
deviations of the density, flow velocity, and temperature are then
put in dimensionless form and given a Fourier representation,
\begin{equation}
\left\{ \delta y_{\alpha }(\mathbf{k},s) \right\} \equiv \int
d\mathbf{r}e^{i\mathbf{ k\cdot r}}\left\{ \frac{\delta n\left(
\mathbf{r},t\right) }{n},\frac{ \delta T\left( \mathbf{r},t\right)
}{T(t)},\frac{\delta \mathbf{u} \left( \mathbf{r},t\right)
}{v_{0}(t)}\right\} .  \label{4.13}
\end{equation}
The resulting linear hydrodynamic equations take the form
\begin{equation}
\partial _{s}\delta y_{\alpha }(\mathbf{k},s)+M_{\alpha \beta }(\mathbf{k})
\delta y_{\beta }(k,s)=0.  \label{4.14}
\end{equation}

The hydrodynamic modes are defined to be the eigenvalues of
$M_{\alpha \beta }(\mathbf{k})$. Interestingly, the expression for
$M_{\alpha \beta }(\mathbf{ k})$ up through order $k$ does not
depend on the assumed forms for the pressure tensor and heat flux;
only the fact that the cooling rate is normal and dimensional
analysis is required. Consequently, the corresponding eigenvalues
can be considered as known exactly up through this order. They are
found to be:
\begin{equation}
\left\{ \lambda_{n}^{(h)}(k) \right\} = \left\{ 0,\frac{\zeta
^{\ast }}{2},-\frac{\zeta ^{\ast }}{2},-\frac{\zeta^{\ast}
}{2},-\frac{\zeta ^{\ast }}{2}\right\} + \mbox{terms of order
$k^{2}$} . \label{4.15}
\end{equation}

\subsection{Existence of Hydrodynamic Modes}

The search for hydrodynamic excitations in the spectrum of the
operator $\left( i\mathbf{k }^{\ast }\mathbf{\cdot v}^{\ast
}+\Lambda^{*} \right) $ can be performed by assuming they re
analytic in $k$ and looking first for the $k=0$ solutions of
(\ref{4.8}). The practical issue of constructing these modes at
finite $k$ is a separate issue from their existence and is
discussed in the next subsection.

Consider again Eq. (\ref{3.13a}) for the HCS solution in terms of
the nonlinear Boltzmann collision operator. Although it is an
equation for the velocity dependence of the distribution function,
it is parameterized by the hydrodynamic fields. Sequential
differentiation with respect to these fields introduces the
linearized Boltzmann collision operator,
\begin{equation}
\frac{\partial }{\partial y_{\alpha }}\frac{\zeta}{2}
\frac{\partial}{\partial \mathbf{v}} \cdot \left[ \left(
\mathbf{v-u}\right) f_{hcs}\right] = \frac{\partial }{\partial
y_{\alpha }}J\left[ \mathbf{v}|f_{hcs} \right] =-f_{hcs}
(\mathbf{v})L\frac{\partial \ln f_{hcs}}{\partial y_{\alpha }}.
\label{4.16}
\end{equation}

This gives five equations relating the operation of $L$ on
derivatives of the HCS solution to other derivatives of the same
distribution. In detail, these terms can be rearranged to
construct eigenvalue equations for $\Lambda^{*}$. In this way five
eigenfunctions and eigenvalues are directly identified
\cite{BDR03,DB03,DB04}
\begin{equation}
\left\{ \lambda_{n}(0) \right\} = \left\{
0,\frac{\zeta^{*}}{2},-\frac{\zeta^{*}}{2},-\frac{\zeta^{*}}{2},
-\frac{\zeta^{*}}{2} \right\}. \label{4.15a}
\end{equation}
\begin{equation} \left\{\psi _{n}(0) \right\} =
\left\{ 4+ \mathbf{v}^{\ast } \cdot \frac{\partial \ln
f_{hcs}^{\ast }}{\partial \mathbf{v}^{*}}  ,-3-\mathbf{v}^{\ast }
\cdot \frac{\partial \ln f_{hcs}^{\ast }}{\partial \mathbf{v}^{*}}
,- \frac{\partial \ln f_{hcs}^{\ast }}{\partial \mathbf{v}^{*}}
\right\} . \label{4.18}
\end{equation}
Clearly, these are the hydrodynamic modes in the long wavelength
limit defined by (\ref{4.15}). This is therefore a direct
confirmation of the existence of hydrodynamic excitations for a
granular gas, independent of other derivations such as the
Chapman-Enskog expansion.

\subsection{Navier-Stokes Order Modes}

Returning to the eigenvalue problem (\ref{4.12}) at finite $k$,
the hydrodynamic modes can be calculated for small $k$ by a
perturbation expansion
\begin{equation}
\lambda _{n}\left( k\right) =\lambda _{n}^{(0)}+k\lambda
_{n}^{(1)}+k^{2}\lambda _{n}^{(2)}+ \cdots, \label{4.19}
\end{equation}
\begin{equation}
\psi _{n}\left( k\right) =\psi _{n}^{(0)}+k\psi
_{n}^{(1)}+k^{2}\psi _{n}^{(2)}+ \cdots, \label{4.19a}
\end{equation}
with $\lambda _{n}^{(0)}=\lambda _{n}(0)$ and $\psi
_{n}^{(0)}=\psi _{n}(0)$ given by (\ref{4.15}) and (\ref{4.18}),
respectively. The analysis is complicated by the three-fold
degeneracy associated with the unperturbed eigenvalues $-\zeta/2$.
These are the eigenvalues of a vector, which can be decomposed
into a longitudinal component along $\widehat{\mathbf{k}}^{*}$ and
two transverse components. The operator $i\mathbf{k}^{\ast
}\mathbf{\cdot v}^{\ast }+\Lambda ^{\ast }$ is invariant under
rotations about $\widehat{\mathbf{k}}^{*}$ so the Hilbert space
can be decomposed into two invariant subspaces comprised of the
rotationally invariant functions and their orthogonal complement.
The transverse and longitudinal eigenfunctions lie in different
subspaces. If $\psi _{3}^{(0)}$ is taken to be the longitudinal
component, the eigenvalue problem can be separated into two
independent eigenvalue problems for $n=1,2,3$ and for $n=4,5$,
respectively. The eigenvalues for the first problem are all
distinct, and therefore standard perturbation theory applies. The
second problem still has a two-fold degeneracy that remains even
at finite $k$.

The fact that $i\mathbf{k}^{\ast }\mathbf{\cdot v}^{\ast }+\Lambda
^{\ast }$ is not Hermitian is a second complication. The
calculations are simplified by introducing a set that is
biorthogonal to $\left\{ \psi _{n}^{(0)}\right\} $,
\begin{equation}
\phi _{n}^{(0)}\rightarrow \left\{
1,\frac{v^{2}}{3}+\frac{1}{2},\widehat{ \mathbf{k}}^{*}\cdot
\mathbf{v}^{*},\widehat{\mathbf{e}}^{(i)}\cdot \mathbf{v}^{*}
\right\} , \quad \quad \left( \phi _{n}^{(0)},\psi
_{m}^{(0)}\right) =\delta _{nm}, \label{4.21}
\end{equation}
where $\left\{ \widehat{\mathbf{k}}^{*},
\widehat{\mathbf{e}}^{(i)}; i=1,2 \right\}$ are three pairwise
orthogonal unit vectors. The determination of the eigenvalues and
eigenvectors to order $k^{2} $ is then straightforward but lengthy
and the details will be given elsewhere \cite{BD05}. Here, only
the results for the eigenvalues are quoted. First, all corrections
to first order in $k$ vanish as expected from the result
(\ref{4.12}). To order $k^{2}$ the coefficients are
\[
\lambda _{1}^{(2)}=\left( \widehat{\mathbf{k}}\mathbf{\cdot
v},\Lambda^{* -1}\widehat{\mathbf{k}}\mathbf{\cdot v}
\psi_{1}^{(0)}(\mathbf{v}) \right),
\]
\[
\lambda _{2}^{(2)}=\left( \widehat{\mathbf{k}}\mathbf{\cdot v}\phi
_{2}^{(0)}(\mathbf{v}),\left( \Lambda^{*} -\frac{\zeta ^{\ast
}}{2} \right) ^{-1}\widehat{ \mathbf{k}}\mathbf{\cdot v}\psi
_{2}^{(0)}\right) + \mathcal{Z}[f^{*}_{hcs}, \psi_{2}^{(0)}],
\]
\[
\lambda _{3}^{(2)}=\left( \left( \widehat{\mathbf{k}}\mathbf{\cdot
v}\right) ^{2}, \left( \Lambda^{*} +\frac{\zeta ^{\ast
}}{2}\right) ^{-1}\widehat{\mathbf{k}}\mathbf{\cdot v}\psi
_{3}^{(0)}(\mathbf{v})\right),
\]
\begin{equation}
\lambda _{4,5}^{(2)}=\left( \left(
\widehat{\mathbf{e}}^{(1,2)}\mathbf{\cdot v}\right) \left(
\widehat{\mathbf{k}}\mathbf{\cdot v}\right) ,\left( \Lambda^{*}
+\frac{\xi ^{\ast }}{2}\right)
^{-1}\widehat{\mathbf{k}}\mathbf{\cdot v} \psi _{4,5}^{(0)}
(\mathbf{v})\right). \label{4.22}
\end{equation}
In the expression of $\lambda_{2}^{(2)}$, $\mathcal{Z}$ is a
rather involved functional of $f^{*}_{hcs}$ and $\psi_{2}^{(0)}$
whose explicit expression will be not given here \cite{DB03}. It
corresponds to the second order in the gradients contributions
that were neglected upon writing expression (\ref{3.7}) for the
cooling rate in the Navier-Stokes approximation. The corresponding
results obtained from the linearized Navier-Stokes equations are
given in terms of the transport coefficients, which are defined
there from the Chapman-Enskog method. It is possible to show that
the results (\ref{4.22}) are exactly the same as those from the
Chapman-Enskog method, and independent confirmation of the latter.

It has been stressed in section 3 that ``hydrodynamics'' should be
understood in a more general context than the Navier-Stokes
approximation. In the present section that means the existence of
modes $\left\{ \lambda _{n}(k)\right\} $ beyond quadratic order in
$k$. For elastic collisions, it has been shown that the
perturbation expansion converges \cite{McL65} so that all higher
order terms are meaningful. However, the proof does not extend to
granular gases. Instead, support for the existence of
hydrodynamics beyond Navier-Stokes order can be obtained from
realistic but simpler kinetic model equations. In reference
\cite{BD04} such a kinetic model is solved exactly (reduced to
quadratures) to demonstrate the hydrodynamic modes in the spectrum
of the general solution to the linear kinetic equation, and to
display the equation for the dispersion relations determining
$\left\{ \lambda _{n}(k)\right\} $. In that case, complete
information about the rest of the spectrum is available as well,
so that the context of a hydrodynamic description can also be
studied.

\section{Summary and Outlook}

The origin of hydrodynamics starts with the balance equations.
These have an exact basis in statistical mechanics and are little
more than a statement of conservation laws at the macroscopic
level. Such equations are correct even for very complex fluids,
solids, composites, or worse. The next step to obtain a
\emph{closed} set of equations for the fields (hydrodynamic,
elastic, viscoelastic, other) is the point of conceptual and
practical difficulty. The choice of constitutive equations is
simple only for systems with well-defined symmetries and response
characteristics: e.g., simple atomic fluids, elastic solids,
liquid crystals. In these cases, the symmetries and time scales
restrict the form of the constitutive equations and the resulting
macroscopic equations have little uncertainty beyond the numerical
values of the constants appearing in them.

Constitutive equations for complex fluids (e.g., polymers,
colloids, foams) are more difficult to fix as neither symmetries
nor characteristic space and time scales are easily determined.
Surprisingly, granular gases are more like complex fluids than
like simple atomic fluids, in spite of their simple structure. To
a great extent this is due to the single feature of inelastic
collisions. Corn starch flows like a simple gas under gravity, but
responds to large shear as a solid. It is probably too ambitious
to attempt to find constitutive equations encompassing both
classes of phenomena. Here attention has been limited to
conditions of low density gas flow, but even with this restriction
it subsumes common conditions of both subsonic and supersonic
flows for granular gases. Thus, it may be expected that simple
constitutive relations may have more limited applicability than
for normal gases.

In the first part of this presentation, the notion of constitutive
equations has been tied to the existence of a normal solution to
the Boltzmann equation. If a normal solution applies, a
hydrodynamic description applies (albeit with possibly complicated
constitutive equations). The idea of a normal solution is
suggested by a two stage process of evolution following
preparation of the system. The first is a rapid velocity
relaxation in each small cell of the gas to a distribution
characterized by the hydrodynamic fields at each point, and also
by the gradients of these fields, reflecting the fact that each
cell is an open system. The second stage is an ``equilibration''
of the spatial variations in the fields from cell to cell, as
enforced by the macroscopic balance equations. Important open
questions at this point include: 1) what is the precise definition
of a normal solution to the Boltzmann equation, i.e. how is the
problem properly posed?,  2) is there a ``universal'' normal
solution approached for a wide class of preparations?, 3) what is
the time scale for approach to a normal solution and is it short
as compared with the hydrodynamic time scales? The Chapman-Enskog
perturbation expansion gives partial answers to these questions in
a very restricted context. Regarding 1), the problem is well-posed
with solutions assured at each order in the perturbation. The
Chapman-Enskog normal solution is indeed a universal function of
the fields and their gradients. Different hydrodynamic states
result only from boundary and initial conditions determining the
values of these fields. Regarding 2) and 3), the Chapman-Enskog
method gives no information.

The second part of this presentation considers the linear response
to small spatial perturbations of a homogeneous reference state.
The question posed is whether the hydrodynamic excitations occur
in the spectrum of the generator for this linear dynamics. The
primary result reported here is that the hydrodynamic excitations
do exist in the long wavelength limit (small $k$). Under the
assumptions of analyticity in $k$ and convergence of a formal
perturbation expansion, the hydrodynamic spectrum is established.
This approach to hydrodynamics for a granular gas provides a
somewhat sharper mathematical formulation of the questions: 1)
what is the appropriate function space to study the spectrum of
$\left( i\mathbf{k}^{\ast }\mathbf{ \cdot v}^{\ast }+\Lambda^{*}
\right) $? and, 2) is the hydrodynamic point spectrum found here
isolated and smaller than all other parts of the spectrum?

Simple experiments on driven systems exhibit complex phenomena
(e.g., symmetry breaking) that nevertheless can be explained by
the correct Navier-Stokes hydrodynamics. Then, complex phenomena
alone should not preclude the simplest Navier-Stokes hydrodynamic
description. On the other hand, some steady state conditions are
necessarily non-Newtonian (beyond Navier-Stokes). This should not
be interpreted as a failure of hydrodynamics, but only as a
challenge to explore the correct constitutive equations for these
conditions.

\bigskip

\section{Acknowledgments}

The research of JWD was supported in part by a grant from the U.
S. Department of Energy, DE-FG02ER54677. The research of J.J.B.
was partially supported by the Ministerio de Ciencia y
Tecnolog\'{\i}a (Spain), through Grant No. BFM2002-00307
(partially financed by FEDER funds).




%





%

%


%


%




\begin{thebibliography}{[1]}

%


%


%







\bibitem[1]{G01} Goldhirsh, I.: Granular gases : Probing the
Boundaries of Hydrodynamics. In: P\"{o}schel, T. and Luding, T.
(eds) Granular Gases. Springer, Berlin Heidelberg New York (2001)

\bibitem[2]{BDR03} Brey, J. J., Dufty, J. W., and  Ruiz-Montero, M. J.:
Linearized Boltzmann Equation and Hydrodynamics for Granular
Gases. In: P\"{o}schel, T. and Brilliantov, N. (eds) Granular Gas
Dynamics. Springer, Berlin Heidelberg New York (2003)


\bibitem[3]{DB03} Dufty, J. W. and Brey, J. J.: Hydrodynamic Modes for a Granular Gas.
Phys. Rev. E \textbf{68}, 030302 (2003)


\bibitem[4]{DB04} Dufty, J. W. and Brey, J. J.: Some Aspects of the Boltzmann
Equation for Granular Gases. In Proceedings of Rarefied Gas
Dynamics 24 (to be published)


\bibitem[5]{BDKS99} Brey, J. J., Dufty, J. W, Kim C. S., and Santos, A.:
Hydrodynamics for Granular Flow at Low Density. Phys. Rev. E
\textbf{58},4638-4653 (1998)


\bibitem[6]{SGD03} Santos, A., Garzo, V., and Dufty, J. W.: Inherent Rheology
of a Granular Fluid in Uniform Shear Flow. Phys. Rev. E
\textbf{69}, 061303 (2004)


\bibitem[7]{BD04} Baskaran, A., and Dufty, J. W.: Hydrodynamics
for a granular gas from an exactly solvable kinetic model. In:
this volume


\bibitem[8]{GS95} Goldshtein A. and Shapiro M.: Mechanics of collisional motion of
granular materials. J. Fluid Mech. \textbf{282}, 75-114 (2004)


\bibitem[9]{D01} Dufty, J. W.: Kinetic Theory and Hydrodynamics for a
Low Density Granular Gas. Advances in Complex Systems \textbf{4},
397-406 (2001)


\bibitem[10]{vNE01} van Noije, T. P. C. and Ernst, M. J.: Kinetic
Theory of Granular Gases. In Poschel, T. and Luding, T. (eds)
Granular Gases. Springer,  Berlin Heidelberg New York (2001)


\bibitem[11]{GD99} Garzo, V. and Dufty, J.: Dense Fluid Transport for Inelastic Hard
Spheres. Phys. Rev. E \textbf{59}, 5895 (1999)


\bibitem[12]{BD05} Brey, J. J., and Dufty, J. W.: Hydrodynamic Modes for a Granular Gas.
Unpublished


\bibitem[13]{McL65} McLennan, J. A.: Convergence of the Chapman-Enskog
Expansion for the Linearized Boltzmann Equation. Phys. Fluids
\textbf{8}, 1580 (1965)


\end{thebibliography}

%




\printindex

\end{document}